\documentclass[letterpaper, 10 pt, conference]{ieeeconf}  % Comment this line out
\IEEEoverridecommandlockouts                              % This command is only
\usepackage[utf8]{inputenc}
\usepackage[T1]{fontenc}

\usepackage{listings}
\usepackage{color}
\usepackage{xcolor}
\usepackage{subcaption,graphicx,dblfloatfix}
\usepackage{hyperref}

\newcommand{\squeezeuphalf}{\vspace{-1.5mm}}

\title{\LARGE \bf
Differentiable Computational Geometry for 2D and 3D machine learning
}

\author{Yuanxin Zhong$^{1}$% <-this % stops a space
\thanks{*This work was not supported by any organization}% <-this % stops a space
\thanks{$^{1}$Yuanxin Zhong is with Mechanical Engineering, University of Michigan, Ann Arbor, USA.
        {\tt\small (zyxin at umich.edu)}}%
}

\begin{document}

\maketitle
\thispagestyle{empty}
\pagestyle{empty}

%%%%%%%%%%%%%%%%%%%%%%%%%%%%%%%%%%%%%%%%%%%%%%%%%%%%%%%%%%%%%%%%%%%%%%%%%%%%%%%%
\begin{abstract}

With the growth of machine learning algorithms with geometry primitives, a high-efficiency library with differentiable geometric operators are desired. We present an optimized Differentiable Geometry Algorithm Library (DGAL) loaded with implementations of differentiable operators for geometric primitives like lines and polygons. The library is a header-only templated C++ library with GPU support. We discuss the internal design of the library and benchmark its performance on some tasks with other implementations.

\end{abstract}

%%%%%%%%%%%%%%%%%%%%%%%%%%%%%%%%%%%%%%%%%%%%%%%%%%%%%%%%%%%%%%%%%%%%%%%%%%%%%%%%
\section{INTRODUCTION}

Geometric representation and computation became a foundation for various computer graphics problems including rendering, collision checking, etc. Recently, computational graphics techniques have been generalized to a wide variety of problems including robotics navigation and object detection. % introduce usage of geometric computation (can be copied from CGAL paper)

There exist various libraries supporting geometric computation including boost.polygon \cite{schaling2011boost}, boost.geometry \cite{gehrels2010generic}, CGAL \cite{fabri2000design}, libigl \cite{jacobson2017libigl}, GTE \cite{GeometricToolsEngine5}, GEOS \cite{libgeos} and even the QT GUI library. They are optimized and have been used in a wide range of applications. Most of these libraries support operations on 2D and 3D geometry primitives, but they are not designed to be used for machine learning applications.

On the other hand, differentiable libraries for geometry related operations have been developed recently, providing the foundation for machine learning with 2D or 3D geometry. Taichi and Kornia\cite{riba2020kornia} are the famous ones among them. Taichi\cite{hu2018taichi} created a reusable infrastructure specifically designed for computer graphics rendering and simulation, its successor DiffTaichi\cite{hu2019difftaichi} created differentiable operators based on that. Kornia is a differentiable computer vision library equipped with differentiable transformation operations widely used in computer vision applications. Although they provide optimized implementations of various geometry operations, they're usually related to transformation and simulation on a rasterized data structure (like pixels or voxels). They lack the support for operations on vectorized geometry primitives like polygons. On the other hand, auto gradient calculation functionality built in many deep learning frameworks like PyTorch\cite{paszke2019pytorch} supports differentiable operations automatically, but since a lot of loops and condition statements are involved in geometry operations like polygon intersection, the efficiency of the automatic gradient calculation is not satisfying. 

To meet the demand for optimized geometry computation, we present the Differentiable Geometry Algorithms Library (DGAL) to support the geometric operations in parallel and with GPU support.

\section{DESIGN GOALS}

\subsection{Accessibility \& Portability}

The DGAL library is an open-source header-only C++ library with CUDA support. No extra dependencies and linking are needed. This also helps to make the library platform-independent. The simple file structure also makes it easy to be integrated into other languages and libraries like Pytorch. The code of our library is available at \url{https://github.com/cmpute/dgal}.

\subsection{Simplicity}
The interfaces defined in the library are very simple and intuitive. Following codes show an example for the intersection over union (IoU) calculation of two quadrilaterals\footnote{For more examples, please refer to the dgal github repository.}.
\begin{lstlisting}[label=iou-code]
#include <dgal/geometry.hpp>
#include <dgal/geometry_grad.hpp>
using namespace dgal;
typedef Poly2<float, 4> Quad2; // define the quadrilateral type

float iou(const Quad2 &p1, const Quad2 &p2, uint8_t &nx, std::array<uint8_t, 8> &xflags)
{ // nx, xflags are intermediate variable for backward calculation
    auto intersection = intersect(p1, p2, xflags.data());
    nx = intersection.nvertices;
    auto area_i = area(intersection);
    auto area_u = area(p1) + area(p2) - area_i;
    return area_i / area_u;
}
std::pair<Quad2, Quad2>
iou_backward(const Quad2 &p1, const Quad2 &p2, const float &grad, const uint8_t nx, const std::array<uint8_t, 8> &xflags)
{ // gradient calculation of IoU is covered by DGAL
    Poly<float, 4> grad1, grad2;
    iou_grad(p1, p2, grad, nx, xflags, grad1, grad2);
    return std::make_tuple(grad1, grad2);
}
\end{lstlisting}

\subsection{Optimized}
All the classes and functions in the library are templated with options for precision and size. The objects like Polygons and Meshes will have variable sizes when performing operations, for instance, the intersection of the two rectangles can have 3-8 vertices. This makes it hard to parallelize the computation. In DGAL we use a fix-size allocated memory for variable-length objects (like C++ vector) to make it easier for parallelization. The memory footprint of geometric operations is also optimized to make the operations sufficiently light-weight to be used in a GPU kernel function.

% TODO: ensure stability?

\section{DATA STRUCTURE}

In this section, we explain how some basic data structures are represented.
\subsubsection{Points} are stored with their 2D or 3D coordinates.
\subsubsection{Lines} are stored with the coefficients of their equation. Lines can be defined by two points in a differentiable manner.
\subsubsection{(Convex) Polygons} are stored in a fixed size point array. The vertices of the polygon cannot be more than the capacity and they are stored in counter-clockwise order.

These data structures are not only used to store the original data but also for storage of the gradients. For example, the Polygon class can also be utilized for containing the gradient value of each of its vertices. Nevertheless, only convex shapes are supported right now. The support of operations for arbitrary polygons is in the plan.

\section{CASE STUDY}

\subsection{Rotated 2D IoU Loss}

\begin{figure}[t]
    \centering
    \begin{subfigure}[b]{\columnwidth}
        \includegraphics[width=\columnwidth]{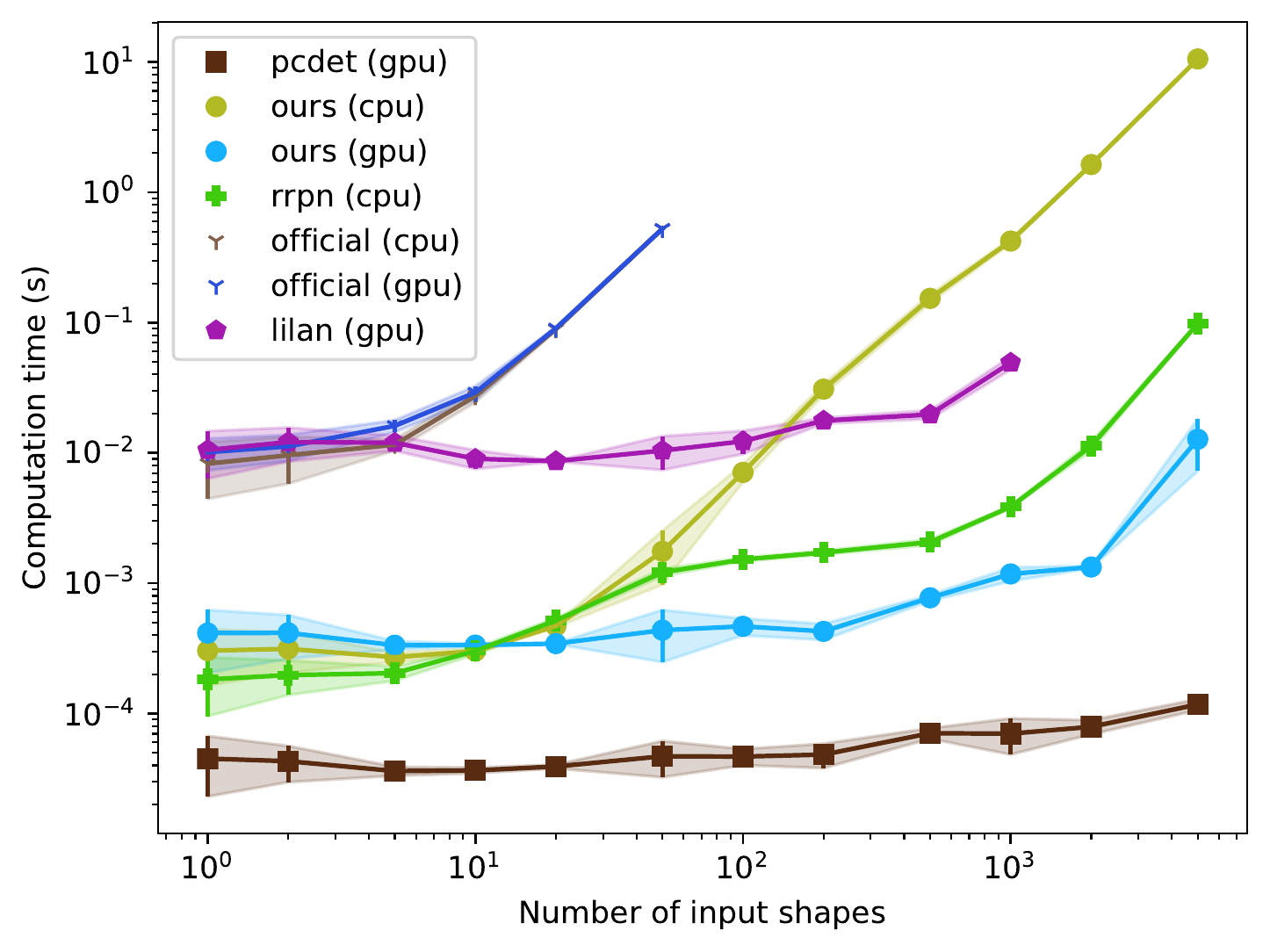}
        \caption{}
    \end{subfigure}
    \hfill
    \begin{subfigure}[b]{\columnwidth}
        \includegraphics[width=\columnwidth]{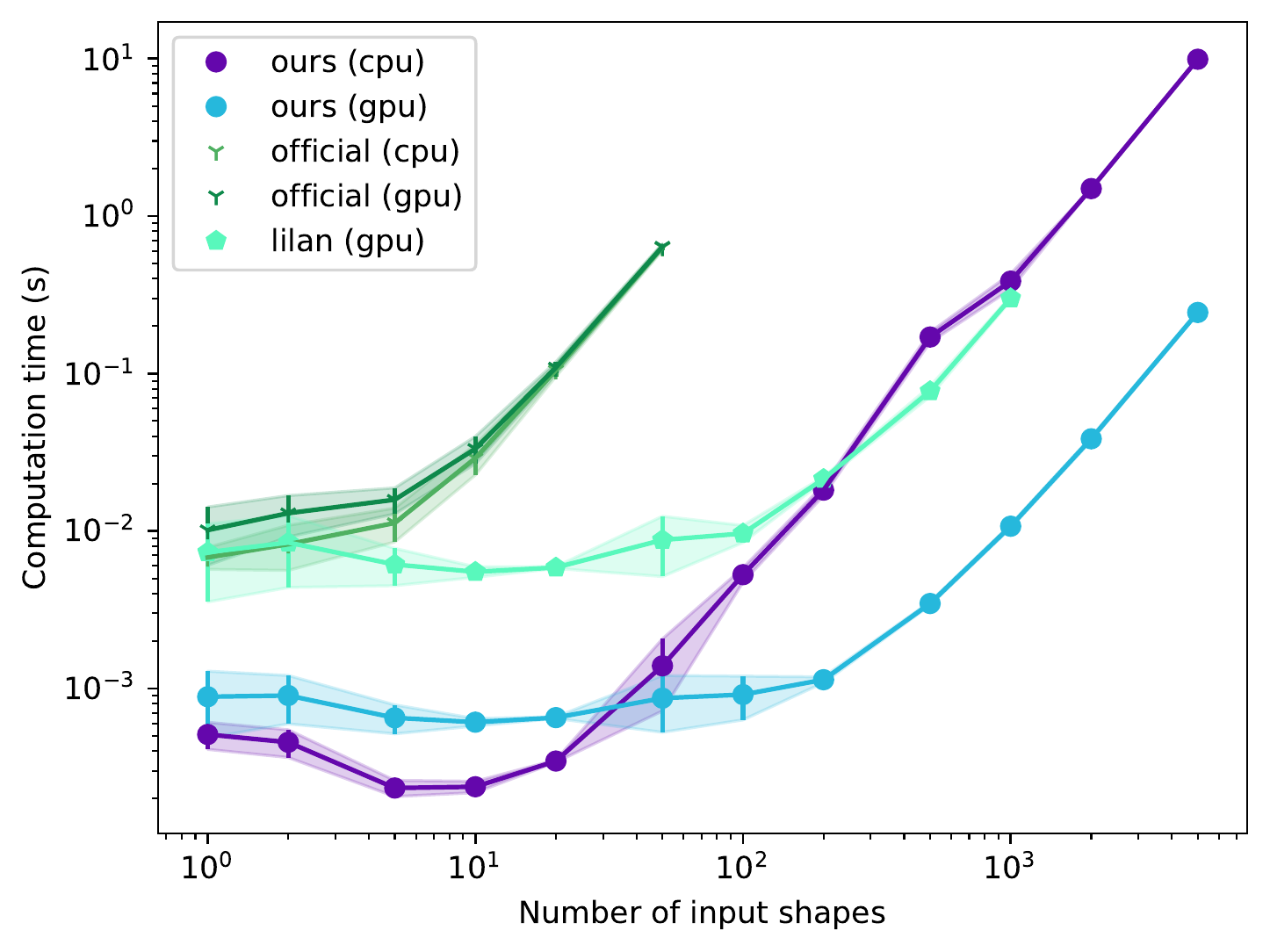}
        \caption{}
    \end{subfigure}
    \caption{Plot (a) and (b) shows the time consumption for forward and backward (gradient) rotated IoU calculations. The performance line labeled as \textit{official} is from the original implementation of \cite{zhou2019iou} and ones labeled as 
    \textit{rrpn}\textsuperscript{a}, \textit{pcdet}\textsuperscript{b}, \textit{lilan}\textsuperscript{c} are third-party implementations.
    }
    Source code: \small\textsuperscript{a}\href{https://github.com/hongzhenwang/RRPN-revise/blob/master/lib/rotation/rbbox.py}{github:hongzhenwang/RRPN-revise}; 
    \small\textsuperscript{b}\href{https://github.com/open-mmlab/OpenPCDet/blob/master/pcdet/ops/iou3d_nms/iou3d_nms_utils.py}{github:open-mmlab/OpenPCDet}; 
    \small\textsuperscript{c}\href{https://github.com/lilanxiao/Rotated_IoU}{github:lilanxiao/Rotated\_IoU}
    \label{fig:performance}
    \squeezeuphalf
\end{figure}

To demonstrate the efficacy of the DGAL library, we present a direct usage of our library in the task of calculating 2D IoU Loss and 3D IoU Loss for rotated bounding boxes (opposite to axis-aligned bounding boxes). We compare our implementation of forward and backward calculation against official\cite{zhou2019iou} and some third-party implementations. The comparison of the performance can be found in Figure \ref{fig:performance}, which indicates that our method has a faster computation speed compared with many other methods. The only method being more efficient than ours is the one implemented in the OpenPCDet, which is optimized for only rectangle intersection and doesn't support gradient calculation. The benchmark code can be found in the \textit{d3d} library\footnote{See \href{https://github.com/cmpute/d3d/blob/master/test/compare/benchmark_riou.py}{github:cmpute/d3d/test/compare/benchmark\_riou.py}}.

\section{FUTURE WORK}

The DGAL library currently only supports a limited collection of vectorized geometric operations, which covers only some mostly used 2D operations. In the future we would like to enhance the library making it a differentiable analogy to the more complex and complete CGAL toolset! We also plan to improve the numerical stability of our library to support arbitrary shape input.

\addtolength{\textheight}{-12cm}   % This command serves to balance the column lengths
                                  % on the last page of the document manually. It shortens
                                  % the textheight of the last page by a suitable amount.
                                  % This command does not take effect until the next page
                                  % so it should come on the page before the last. Make
                                  % sure that you do not shorten the textheight too much.

%%%%%%%%%%%%%%%%%%%%%%%%%%%%%%%%%%%%%%%%%%%%%%%%%%%%%%%%%%%%%%%%%%%%%%%%%%%%%%%%

\bibliographystyle{IEEEtran}
\bibliography{ref}

\end{document}